\documentclass[a4paper,twocolumn,prb,amssymb,aps, longbibliography]{revtex4-2}

\usepackage{dcolumn}
\usepackage{bm}
\usepackage{graphicx,color}
\usepackage{epstopdf}
\usepackage{xspace}
\usepackage{amsmath}
\usepackage[english]{babel}
\usepackage{verbatim}
\usepackage{hyperref}
\hypersetup{
    colorlinks=true,
    linkcolor=blue,
    filecolor=magenta,      
    urlcolor=cyan,
    citecolor=blue
    }
\usepackage{braket}
\usepackage{booktabs}
\usepackage{ulem} 
\usepackage[version=3]{mhchem}

\newcommand{\ac}{$\alpha_\mathrm{c}$}

\newcommand{\ain}{$\alpha_\mathrm{ab}$}
\newcommand{\lin}{$\lambda_\mathrm{ab}$}

\newcommand{\TC}{$T_\mathrm{C}$}
\newcommand{\tc}{$T_\mathrm{C}$}
\newcommand{\CGT}{Cr$_2$Ge$_2$Te$_6$}

\newcommand{\alimag}{$\alpha_{i,\mathrm{mag}}$}

\begin{document}
	
	\title{Strong uniaxial pressure dependencies evidencing spin-lattice coupling and spin fluctuations in Cr$_2$Ge$_2$Te$_6$}
	\author{S. Spachmann$^{1}$}
	\author{S. Selter$^{2}$}
	\author{B. B\"uchner$^{2,3}$}
	\author{S. Aswartham$^{2}$}
	\author{R. Klingeler$^{1}$}

	\affiliation{$^1$Kirchhoff Institute for Physics, Heidelberg University, Heidelberg, Germany}
	\affiliation{$^2$Leibniz Institute for Solid State and Materials Research (IFW), Dresden, Germany}
	\affiliation{$^3$Würzburg-Dresden Cluster of Excellence ct.qmat}
	
\begin{abstract}
Single crystals of Cr$_2$Ge$_2$Te$_6$ were studied by high-resolution capacitance dilatometry to obtain in-plane ($B\parallel ab$) and out-of-plane ($B\parallel c$)
thermal expansion and magnetostriction at temperatures between 2 and 300~K and in magnetic fields up to 15~T. The anomalies in both response functions lead to the 'magnetoelastic' phase diagrams and separate the paramagnetic (PM), ferromagnetic low-temperature/low-field (LTF) and aligned ferromagnetic (FM) phases. The presence of two distinct thermal expansion anomalies at small fields $B\parallel ab$ of different magnetic field dependence clearly supports the scenario of an intermediate region separating PM and LTF phases and is indicative of a tricritical point. Simulations of the magnetostriction using the Stoner-Wohlfarth model for uniaxial anisotropy demonstrate that the observed quadratic-in-field behavior in the LTF phase is in line with a rotation of the spins from  the preferred $c$ direction into the $ab$ plane. Both the LTF and the PM phase close to \TC\ exhibit very strong pressure dependencies of the magnetization, ${\partial}\ln{M_{\rm ab}}/{\partial}p_{\rm ab}$, of several hundred \%/GPa and also the transition from the LTF to the FM phase strongly depends on $p_{\rm ab}$ ($\sim -280$\%/GPa), indicating a strong decrease of the uniaxial anisotropy under applied in-plane pressure. Our data clearly demonstrate the relevance of critical fluctuations and magnetoelastic coupling in Cr$_2$Ge$_2$Te$_6$. 

\end{abstract}

\date{\today}
\maketitle

\section{Introduction}
Magnetic anisotropy is one of the key features of quasi-two-dimensional (quasi-2D) magnetic van der Waals (vdW) materials~\cite{Gong2019Review}. Purposefully engineering this anisotropy, e.g. by strain~\cite{Webster2018, Wang2020, Miao2021}, pressure~\cite{Valenta2021, Li2022, Peng2022}, doping~\cite{Park2020, Lee2021} or the combination with other materials in heterostructures~\cite{Cui2020, Dong2020PRB, Ye2021}, is a crucial ingredient in the hunt for 
room-temperature 2D ferromagnets on one end of the spectrum between applied and fundamental research and for (Kitaev) Quantum Spin Liquids (QSLs) on the other end~\cite{Kasahara2018, Xu2020}.
Achieving room-temperature 2D ferromagnetism in vdW materials is desirable for a large number of applications, among them spintronic, magnonic, and spin-orbitronic devices~\cite{Gong2019Review}.
Identifying and engineering (Kitaev) QSLs on the other hand, on top of the significance for fundamental physics, comes with the promise of topologically-protected quantum computing~\cite{Freedman2003} due to presence of Majorana Fermions~\cite{Banerjee2018}.
In the quests to engineer the properties and especially the anisotropy of magnetic quasi-2D materials, the interlayer and intralayer spacings play a significant role. Only small changes of these spacings can lead to drastic responses in magnetic and electronic properties~\cite{Blei2021}.
These drastic responses originate from the delicate interplay of charge, spin, and lattice degrees of freedom in correlated electron systems.

The transition metal trichalcogenide \CGT\ is one prominent example of the currently highly-investigated layered magnetic vdW materials.
Its layers, stacked along the $c$ axis, consist of a honeycomb network of edge-sharing \ce{CrTe6} octahedra. \CGT\ crystallizes in the trigonal space group $R\bar{3}$ (No.~148) and becomes ferromagnetic below \TC~$\approx 65$~K in its bulk form. This ferromagnetism is preserved when the material is thinned to a bilayer where long-range magnetic order is observed below about 30~K~\cite{Gong2017}. 

While at ambient pressure \CGT\ exhibits uniaxial anisotropy, conflicting claims about a possible change from uniaxial to easy-plane anisotropy under hydrostatic pressure above 1~GPa have been made~\cite{Lin2018, Sakurai2020}. Importantly, strong spin-lattice coupling has been proven by Raman scattering at ambient and non-ambient pressures~\cite{Tian2016, Sun2018ApplPhysLett}. Furthermore, zero-field thermal expansion measurements of \CGT\ have revealed sharp anomalies both along the in-plane and out-of-plane directions associated with the evolution of long-range magnetic order at \tc , evidencing the presence of significant magnetoelastic coupling~\cite{SpachmannCGT2022}. 

In this paper we investigate the effect of magnetic fields on the structural and magnetic properties of \CGT . In particular, we investigate the effects on the in-plane and out-of-plane lattice parameters with a special focus on the low-temperature and low-field (LTF) phase which is present when a small magnetic field is applied along the in-plane direction. The magnetostriction and magnetization below \TC\ for $B\parallel ab$ are modeled using the Stoner-Wohlfarth model. We establish the magneto-elastic phase diagram and derive and discuss the uniaxial pressure dependencies of the critical field and of the magnetization.

\section{Experimental Details}
Single crystals of \CGT\ have been grown by the self-flux technique and were characterized in detail as reported in Refs.~\onlinecite{Zeisner2019,Selter2020}. The crystals measured in this study have a very thin cuboid shape with a length of 0.255~mm along the $c$ axis for one crystal, as well as in-plane dimensions of $1.3\times 2.0$~mm$^2$ for a second crystal used for in-plane measurements. High-resolution dilatometry measurements were performed by means of a three-terminal high-resolution capacitance dilatometer from Kuechler Innovative Measurement Technology in a home-built set-up~\cite{Kuechler2012, Werner2017}. The capacitance read-off was facilitated by Andeen-Hagerling's AH~2550A Ultra-Precision 1~kHz capacitance bridge.
With the dilatometer, the uniaxial relative length changes ${\Delta}L_i(T)/L_i$ and the linear thermal expansion coefficients $\alpha_i=1/L_i\times dL_i(T)/dT$ both along the $c$ axis and along the in-plane direction, i.e., $\parallel ab$, were studied at temperatures between 2 and 300~K in zero field and in magnetic fields up to 15~T. Magnetic fields were applied along the direction of the measured length changes $i = c$, $\parallel ab$.
In addition, the field-induced length changes ${\Delta}L_i(B_i)$ were measured at various fixed temperatures between 2 and 204~K in magnetic fields up to 15~T. 
Magnetization measurements were performed in a Magnetic Properties Measurement System (MPMS-3) by Quantum Design using the SQUID-VSM option.

\section{Results}


\subsection{Thermal Expansion and Magnetostriction}
\begin{figure*}[htb]
	\center{\includegraphics [width=1.85\columnwidth,clip]{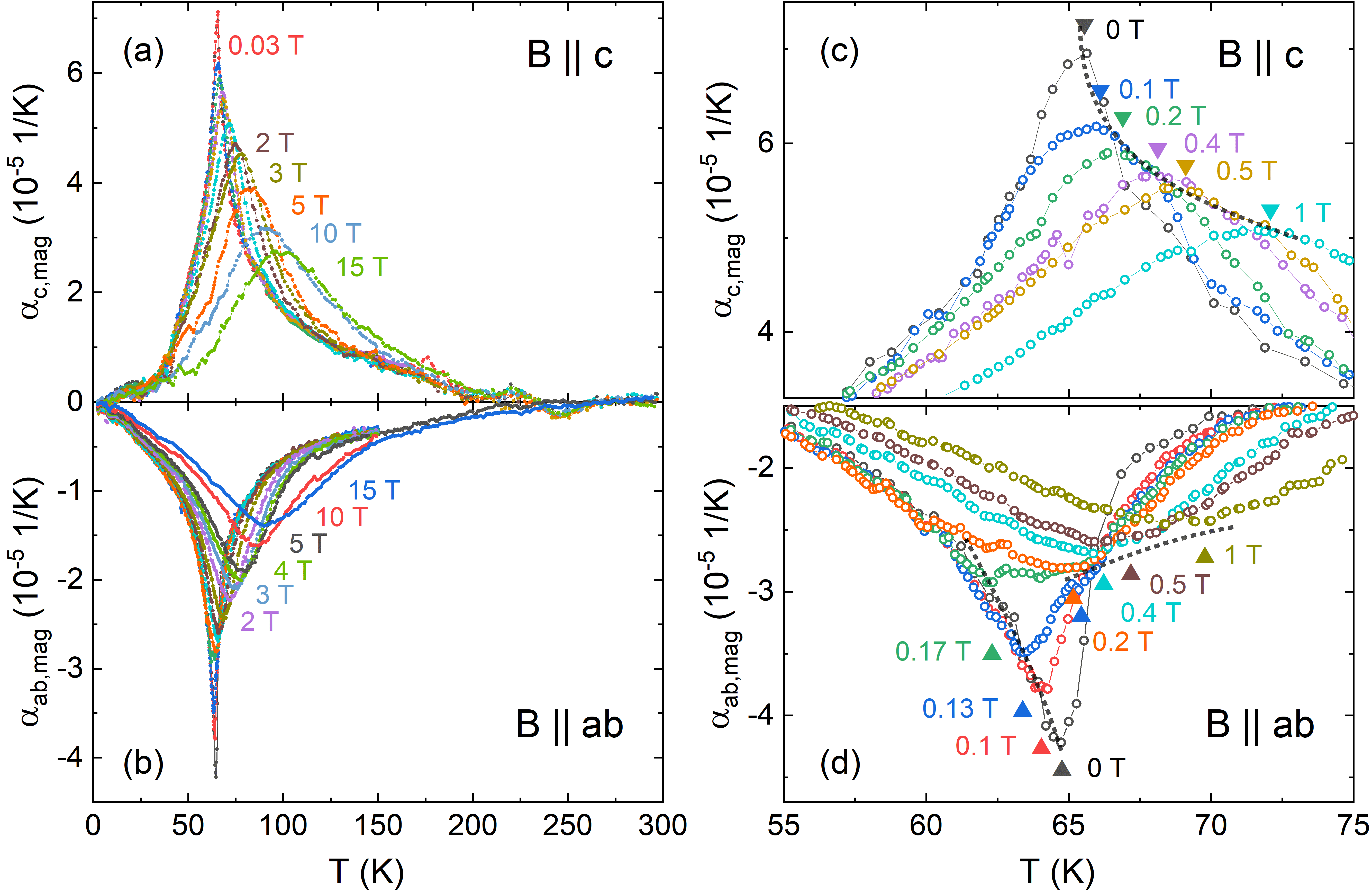}}
	\caption[] {\label{Alpha_mag} Magnetic contributions to the thermal expansion coefficient in external fields between 0 to 15~T (a, b) and between 0 to 1~T (c, d) for $B\parallel c$ and $B\parallel ab$, respectively. Dashed lines in (c, d) indicate trends of the shift of the peak positions, triangles mark the peak positions themselves.}
\end{figure*}

Fig.~1 shows the effect of magnetic field applied along the $c$ axis ($ab$ plane) on the thermal expansion coefficient \ac~(\ain). The same phonon background as in Ref.~\onlinecite{SpachmannCGT2022} was subtracted to obtain the non-phononic contributions \alimag\ shown in the figure. At zero field, pronounced anomalies in the thermal expansion coefficients imply significant magnetoelastic coupling and also confirm the presence of strong spin fluctuations at \TC . As a general trend for both measurement directions, the anomalies in \alimag\ around \TC\ broaden and shift to higher temperatures as the magnetic field is increased.
At small magnetic fields $B\parallel ab$ below 0.2~T, however, an initial suppression of the peak position is observed.
The suppression of the anomaly associated with the onset of the LTF phase implies the presence of a competing ferromagnetic phase which is stabilized in magnetic fields with respect to the LTF phase (while the crossover temperature into the PM phase increases as expected for a ferromagnetic transition). Indeed, the data at 0.13~T show the presence of two anomalies (Fig.~\ref{Alpha_mag}(d) and Fig.~S2 in the supplement). While the lower-in-temperature anomaly confirms the initial suppression of the dominating peak to 63.3~K, it is accompanied by a small peak at 65.4~K (i.e., slightly above \TC ). At 0.17~T, the whole feature further broadens and extends to higher temperature while only a small additional feature remains at 62.3~K. As the field is increased above 0.17~T the broad anomaly continues to shift to higher temperatures whereas the small one cannot be distinguished anymore. 
The observation  of two distinct thermal expansion anomalies in small fields $B\parallel ab$ further confirms the presence of an intermediate phase separating the LTF and PM phases at this field in agreement to Ref.~\cite{Selter2020}. The data indicate that the two anomalies merge to a single point above or at $B=0$~T, giving rise to a tricritical point as suggested in Ref.~\cite{Liu2017}.

\begin{figure}[ht]
	\center{\includegraphics [width=0.9\columnwidth,clip]{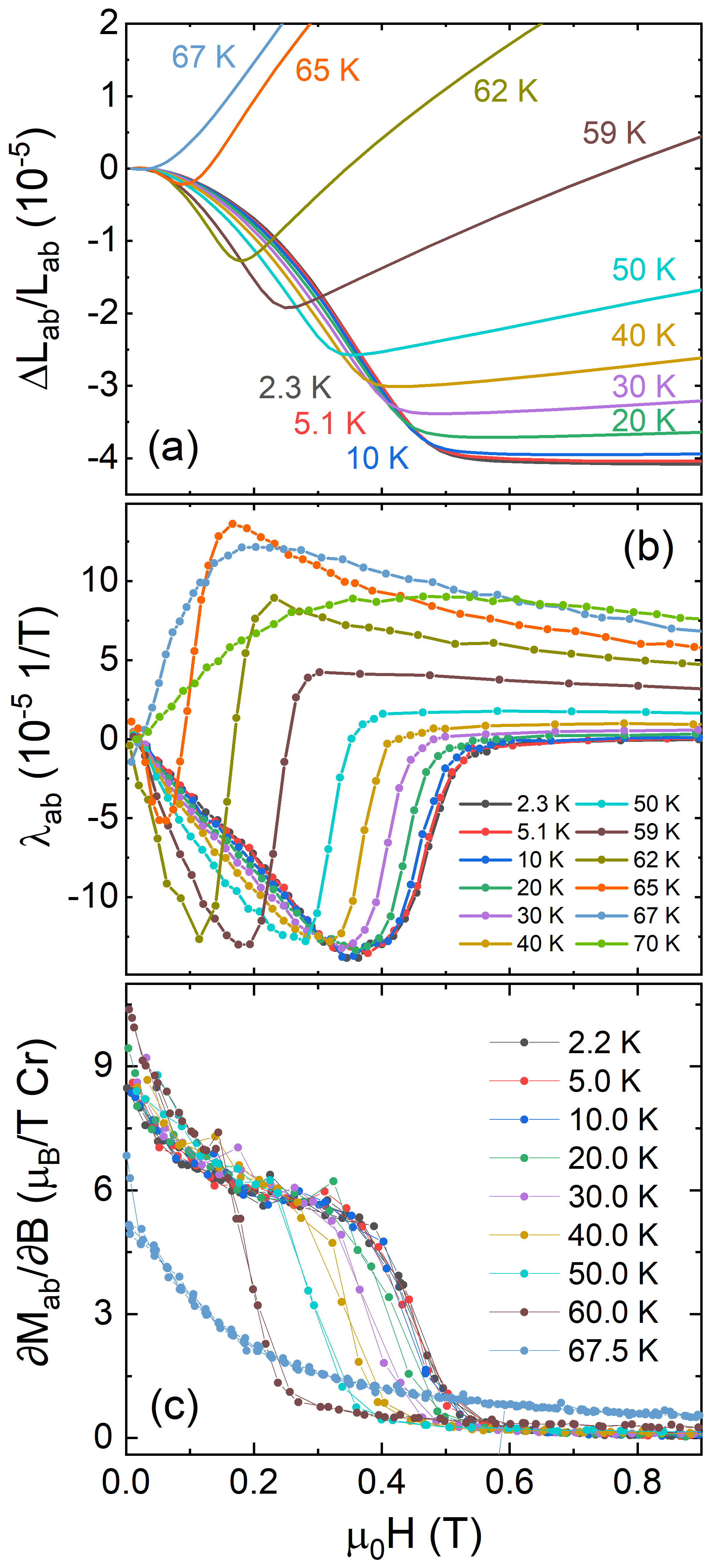}}
	\caption[] {\label{MSMB_In-plane} Relative length changes (a), magnetostriction coefficient (b), and magnetic susceptibility (c) at low fields for $B\parallel ab$ and temperatures from 2 to 70~K. Only up-sweep data are shown in (a) and (b), whereas up- and down-sweeps are shown in (c).}
\end{figure}

In-plane magnetostriction measurements, with $B\parallel ab$, reveal a behavior which clearly implies the presence of two distinct phases below \TC~(Fig.~\ref{MSMB_In-plane}(a) and (b)). 
At low fields and temperatures the relative length changes ${\Delta}L_{\mathrm{ab}}(B)$ exhibit a roughly quadratic-in-field decrease, ${\Delta}L_{\mathrm{ab}}(B) \propto B^2$, before transitioning to a constant value, i.e., vanishing magnetostriction.
Increasing the temperature towards \TC~decreases the range of the quadratic-in-field behavior and the magnetostriction coefficient above the transition assumes a positive value.
The jump in \lin\ suggests that the transition between the two regimes may be interpreted as a second order phase transition.

In-plane measurements of the isothermal magnetization confirm the observed behavior. The magnetic susceptibility  ${\partial}M_{\mathrm{ab}}/{\partial}B$  exhibits a jump at the same fields as \lin~(Fig.~\ref{MSMB_In-plane}(c)).
In contrast, magnetostriction and magnetization measurements with $B\parallel c$ below \TC\ show no phase transition (see supplemental material, Fig.~S4).

\subsection{Simulations Using the Stoner-Wohlfarth Model}
In order to model the low-temperature magnetization and magnetostriction for $B \parallel ab$ the Stoner-Wohlfarth (SW) model~\cite{Stoner1948} was applied.
The normalized energy density in this model -- with $\phi$ being the angle between the magnetic moment, expressed by the saturation magnetization $M_{S}$, and external magnetic field $H$, and $\theta$ being the angle between the easy axis and $H$ -- is given by
\begin{equation}
    \eta = \frac{E}{2K_U V} = \frac{1}{4} -\frac{1}{4} \cos{2(\phi-\theta)}-h \cos{\phi} 
\end{equation}
where the first two terms describe the anisotropy energy density and the third term the Zeeman energy density. The normalized field is given by $h = \mu_{0}M_{S}H/(2K_U)$, where $\mu_{0}$ is the vacuum permeability, $M_S$ the saturation magnetization, and $K_U$ the energy density characterizing the uniaxial anisotropy along $c$.
Since the field was applied in the $ab$ plane, i.e. $\perp c$, we have $\theta = \pi/2$.
$\eta$ was minimized with respect to $\phi$ to obtain the direction of the magnetic moment at a given field. Due to the uniaxial symmetry of the system two values $\phi_{i}$ minimize $\eta$ for $H < H_{\mathrm{sat}}$.
The magnetization along the applied field direction is then obtained as
\begin{align}
    M(h,T) &= \sum_{i=1}^{2} M_S(T) \cos{(\phi_i(h))}\sin{\theta_i}/2 \\
     &=  \sum_{i=1}^{2} M_S(T) \cos{(\phi_i(h))}/2
\end{align}
For $H \geq H_{\mathrm{sat}}$ the magnetic moment is aligned along the field, i.e., $\phi = 0$ and $M(h,T) = M_S(T)$.
$M_S(T)$ is extracted from the magnetization at the intersection of linear fits to the $M(B)$ data at the respective temperature in the field range from 0 to 0.2 T and 3 to 7 T (at 60 K: 0 to 0.15 T and 4.5 to 7 T).
Analogously to the magnetization, the magnetostriction is given by
\begin{align}
    {\Delta}L(h,T) &= \frac{K_{U,\mathrm{eff}}(T)}{K_{U}}\lambda_{\mathrm{sat}} \sum_{i=1}^{2} (\cos{(\phi_i(h))}\sin{\theta_i})^2/2 \\
     &= \frac{K_{U,\mathrm{eff}}(T)}{K_{U}}\lambda_{\mathrm{sat}} \sum_{i=1}^{2} (\cos{(\phi_i(h))})^2/2,
\end{align}
where $\lambda_{\mathrm{sat}}$ is the saturation magnetostriction along the measurement direction and $K_{U,\mathrm{eff}}$ is an effective anisotropy parameter.
This effective anisotropy parameter is calculated according to the theory by Callen and Callen~\cite{Callen1966} as
\begin{equation}\label{eq:Callen-Callen}
\frac{K_{U,\mathrm{eff}}(T)}{K_{U}} = \left[ \frac{M_S(T)}{M_S} \right]^{l(l+1)/2}.
\end{equation}
In the following, we approximate $K_{U} \approx K_{U,\mathrm{eff}}$(2.2~K) and $M_S \approx M_S$(2.2~K).

\begin{figure}[ht]
	\center{\includegraphics [width=0.95\columnwidth,clip]{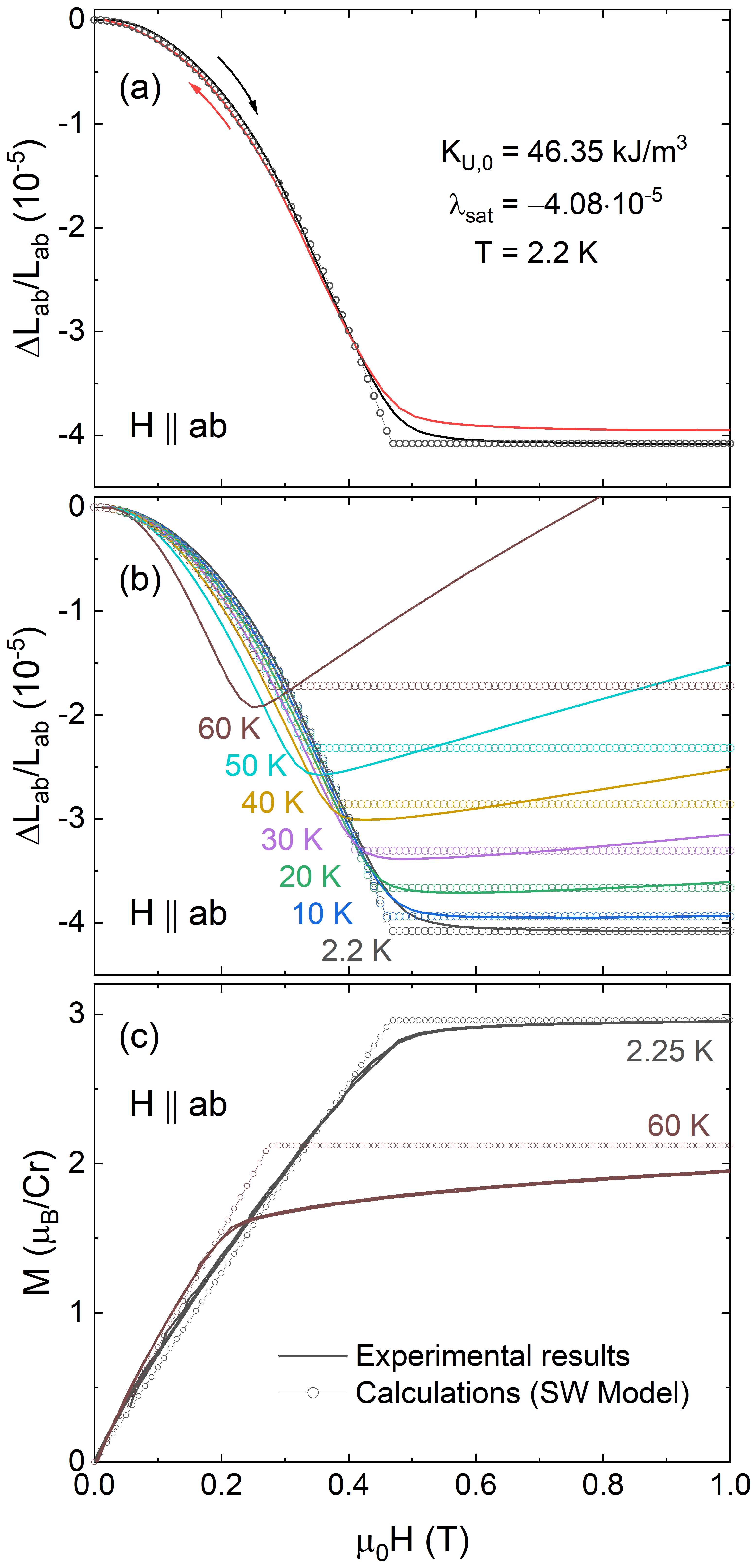}}
	\caption[] {\label{MS_SW-Model} Comparison of experimental magnetostriction data (a, b) and magnetization data (c) for $H \parallel ab$ (solid lines) with calculations using the Stoner-Wohlfarth model (open circles). Only up-sweep data are shown in (b) whereas both up- and down-sweep are shown in (a) and (c).}
\end{figure}

Applying the model as described above, the best agreement between magnetostriction results at 2.2~K (solid line) and the simulation (open circles) is obtained for $K_{U} = 46.35$~kJ/m$^3$ and $\lambda_{\mathrm{sat}} = -4.08\cdot 10^{-5}$ (Fig.~\ref{MS_SW-Model}(a)).
Small deviations between experimental results and simulations are visible at low fields, where the down-sweep data is closer to the simulated values. Also, a smooth transition is observed in the experiments signaling the full alignment of the magnetic moments with the magnetic field at roughly 0.5~T in contrast to a kink in the numerical results. Overall the data are described very well by the simulation.
The same holds for the magnetization at low temperatures (Fig.~\ref{MS_SW-Model}(c)).

Increasing the temperature towards \TC, the experimental data (solid lines) can be described reasonably well using $l = 1.83$ for the calculations (open circles), which is close to $l = 2$ ($n = l(l+1)/2 = 3$) expected for uniaxial anisotropy.
Up to 20~K and 1~T these parameters describe the experimental results well, however, above this temperature and field two types of strong deviations can be distinguished: (1) the experimental magnetostriction is larger at low fields than the simulated one, and (2) a change from zero to positive magnetostriction above the saturation field becomes visible which gets larger in magnitude as the temperature approaches \TC\ (see Fig.~\ref{MS_SW-Model}(b)).
The latter effect is most likely caused by a disturbance of the spin alignment due to phonons and spin-phonon coupling as the temperature is increased.
This explanation is supported by the experimental magnetization which is significantly lower than the one calculated within the SW model (Fig.~\ref{MS_SW-Model}(c)), as well as the previous observation of a significant spin-phonon coupling in \CGT~\cite{Tian2016}.
Note that using an effective anisotropy parameter $K_{U,\mathrm{eff}}$ does not fully cover the effects induced by raising the temperature.

\section{Discussion}

\subsection{Low-Temperature and Low-Field Phase}
The simulations described above show that the magnetostriction and magnetization of the low-temperature and low-field (LTF) phase in \CGT\ can largely be described by the Stoner-Wohlfarth (SW) model at low temperatures. 
This result suggests that a magnetic field applied along the in-plane direction in the LTF phase leads to a rotation of the magnetic moments from the anisotropy-preferred $c$ direction towards the magnetic field until saturation, i.e., full alignment, is reached.
Such a rotation of the magnetic moments in the LTF phase of \CGT\ was first described by S. Selter et al.~\cite{Selter2020}. We also note that the SW model was also used to model ferromagnetic resonance (FMR) data~\cite{Khan2019, Zeisner2019}.
These previous reports obtained anisotropy parameters $K_U = 47(1)$~kJ/m$^{3}$~\cite{Selter2020}, 48~kJ/m$^{3}$~\cite{Zeisner2019} and 39.5~kJ/m$^{3}$~\cite{Khan2019}.
The first two are in perfect agreement with $K_U = 46.35$~kJ/m$^{3}$ found from our simulations.






\subsection{Uniaxial Pressure Dependence of the Critical Field}
For continuous phase transitions the uniaxial pressure dependence of the critical field is given as
\begin{equation}
    \frac{{\partial}B_{\mathrm{cr}}}{{\partial}p_{ab}} = \frac{{\Delta}\lambda}{{\Delta}\left(\frac{{\partial}M}{{\partial}B}\right)}
\end{equation}
where the two quantities in the numerator and denominator on the right-hand side are jumps in the magnetostriction coefficient and in the magnetic susceptibility at the phase boundary, respectively.
For \CGT\ these jumps amount to $\Delta\lambda_{ab} = 1.9(3)\cdot 10^{-4}$/T and $\Delta$(${\partial}M_{ab}/{\partial}B) = -4.6(5)$~$\mu_B$T$^{-1}$Cr$^{-1}$ at 2.2~K, which, using the molar volume of $V_{\mathrm{m}} = 1.67\cdot 10^{-4}$~m$^{3}$/mol, leads to ${\partial}B_{\mathrm{cr}}/{\partial}p_{ab} = -1.25(18)$~T/GPa ($\simeq -280$~\%/GPa). Up to \TC\ this value remains constant within error bars (see supplement, Table 1). Our data hence imply that $B_{\rm cr}$ is extremely sensitive to in-plane pressure. Considering the rather small critical fields of less than 0.5~T, this result implies that the LTF phase can be fully suppressed at small uniaxial pressures, so that the field-induced high-field phase will appear already at $B=0$~T.

This behavior suggests a strong decrease of the uniaxial anisotropy parameter under pressure $p \parallel ab$.
A comparable decrease of the magnetic anisotropy energy (MAE) by compressive biaxial (in-plane) strain has been calculated for a heterostructure of \CGT~\cite{Gao2021} and also for a similar van der Waals material Fe$_3$GeTe$_2$~\cite{Zhuang2016}.
We also note that switching from uniaxial to easy-plane anisotropy has been reported in \CGT\ for hydrostatic pressures larger than 1~GPa~\cite{Lin2018}, which was, however, not confirmed at such low pressures by another study~\cite{Sakurai2020}.
In contrast to uniaxial pressure applied in the $ab$ plane, uniaxial pressure $p \parallel c$ has been shown to strongly stabilize the ferromagnetic (FM) low-temperature phase at the cost of the high-temperature paramagnetic one~\cite{SpachmannCGT2022}, suggesting an enhancement of the uniaxial anisotropy along $c$.

\subsection{Uniaxial Pressure Dependence of the Magnetization}

By exploiting Maxwell relations, the experimentally obtained magnetostriction coefficients $\lambda_i$ can be used to obtain the uniaxial pressure dependencies of the magnetization, i.e.,   
\begin{equation}
    \lambda_{ij} = \frac{1}{L_i}\left(\frac{{\partial}L_i}{{\partial}B_j} \right) 
    = \frac{1}{L_i}\left(\frac{\partial^2 G}{{\partial}p_{i}{\partial}B_j} \right) 
    = -\frac{{\partial}M_i}{{\partial}p_j}
\end{equation}
where $G$ is the Gibbs potential, such that the relative pressure dependence of the magnetization can be calculated as
\begin{equation}
    \frac{\lambda_{ij}}{M_i} = -\frac{{\partial}\ln(M_i)}{{\partial}p_j}. \label{eq2} 
\end{equation}
In the case at hand $i = j$ , i.e., the magnetic field is applied along the direction of the measured length changes.

\begin{figure}[ht]
	\center{\includegraphics [width=1\columnwidth,clip]{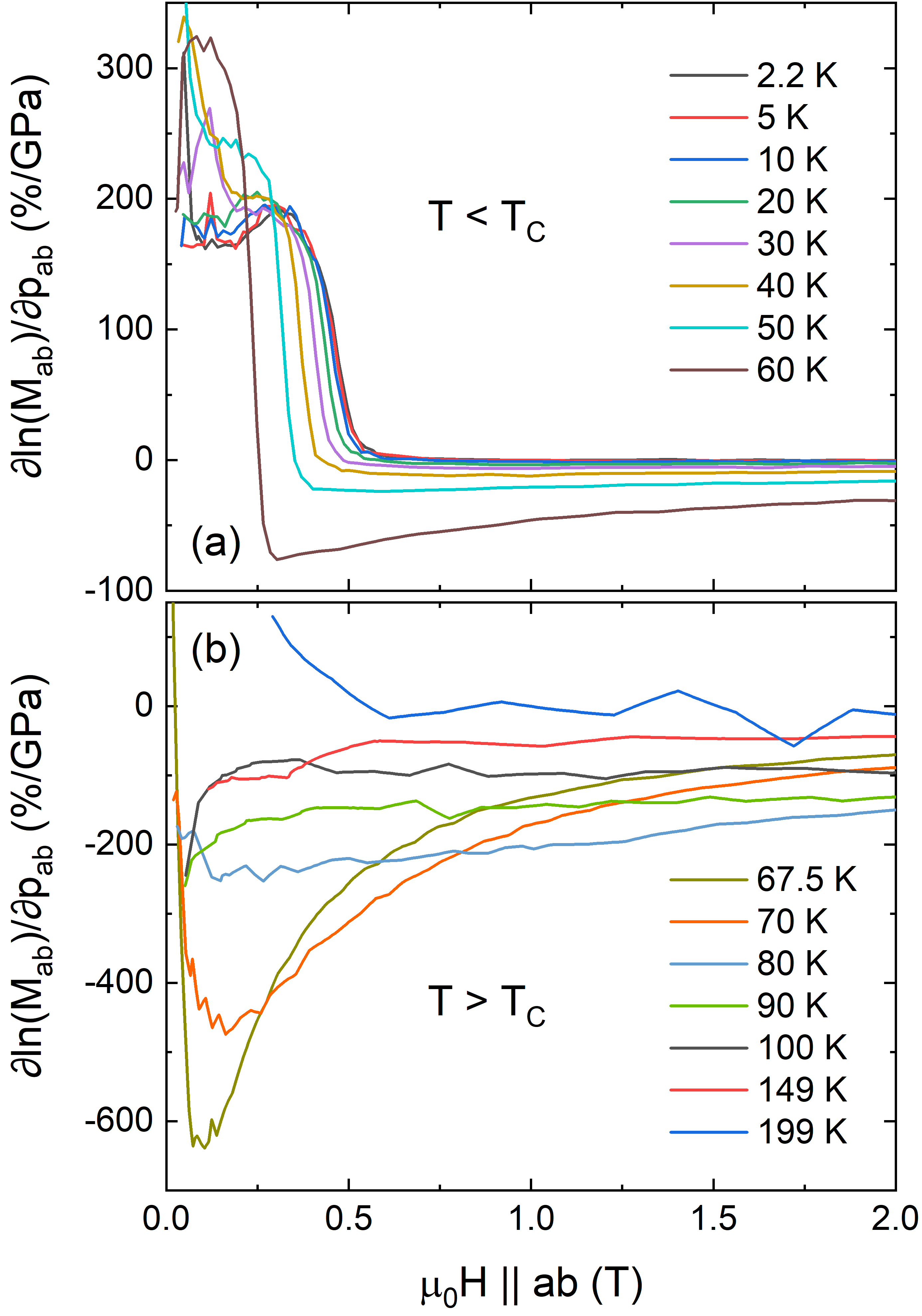}}
	\caption[] {\label{dM-dp} Pressure dependence of the magnetization for $B, p\parallel ab$ at temperatures $T < T_C$ (a) and $T > T_C$ (b).}
\end{figure}

At $T<T_{\mathrm{C}}$, the in-plane magnetostriction exhibits a positive jump in \lin\ at critical fields $B_{\mathrm{cr}}\approx 0.5$~T (see Fig.~2) which according to Eq.~\eqref{eq2} translates to a negative jump in ${\partial}\ln(M_{ab})/{\partial}p_{ab}$. As seen in Fig.~\ref{dM-dp}(a), the uniaxial pressure dependence of the magnetization along $ab$ is large for $B<B_{\mathrm{cr}}$, i.e., $M_{ab}$ will be strongly enhanced by uniaxial pressure $p\parallel ab$. In contrast, for $B>B_{\mathrm{cr}}$ the pressure effect is close to zero at low temperatures and negative with increasing magnitude as the temperature approaches \TC.
Above \TC , ${\partial}\ln(M_{ab})/{\partial}p_{ab}$ is strongly negative with a peak at low fields reaching up to $-600$\%/GPa at 67.5~K (Fig.~\ref{dM-dp}(b)). As the temperature is increased further above \TC\ the pressure dependence of the magnetization decreases until it reaches roughly zero at 200~K.

For $B,p\parallel c$, the uniaxial pressure dependence of the magnetization is positive both in the paramagnetic and the ferromagnetic phase (data not shown). Except for the opposite sign, similar behavior is found as observed for $B,p\parallel ab$, i.e., there is a strong pressure effect of several hundred \%/GPa just above \TC\ at low fields whereas the pressure effect amounts to only a few \%/GPa at $T<50$~K and well above \TC .

Neither measurements of the in-plane magnetization of \CGT\ under (hydrostatic or uniaxial) pressure nor any measurements at all under uniaxial pressure have been reported in the literature. However, the $hydrostatic$ pressure dependence of the magnetization has been measured for $B\parallel c$. Measurements by Sakurai et al.~at 0.1~T show a pressure dependence of the magnetization of about $-20$\%/GPa for pressure up to 3.4~GPa at low temperatures~\cite{Sakurai2020}. In contrast, Bhoi et al.~found only small changes of a few percent up to 1.73~GPa under the same conditions whereas a much stronger pressure dependence also up to about $-20$\%/GPa has been reported at 4~GPa~\cite{Bhoi2021}. While in vdW materials it may be not straightforward to trace back experimentally observed hydrostatic effects to uniaxial ones, these results suggest that under hydrostatic pressure the resulting strain in the $ab$ plane, which leads to the rotation of the magnetic moments into the $ab$ plane, has a stronger effect than the $c$ axis stress. This result is in line with our observation that ${\partial}\ln(M_{ab})/{\partial}p_{ab} \gg {\partial}\ln(M_{c})/{\partial}p_{c}$.
The here reported uniaxial pressure dependencies are much larger than the hydrostatic ones from the literature which is typical in solids. It indicates  mutual canceling of the uniaxial in-plane and out-of-plane pressure dependencies and is in accordance with the cancellation of uniaxial pressure dependencies of the critical temperature in \CGT~\cite{SpachmannCGT2022}. The observed strong uniaxial pressure dependencies around \TC, however, further confirm the relevance and enhancement of critical fluctuations in this regime of the phase diagram. 



\subsection{Phase Diagrams}
\begin{figure}[ht]
	\center{\includegraphics [width=1.0\columnwidth,clip]{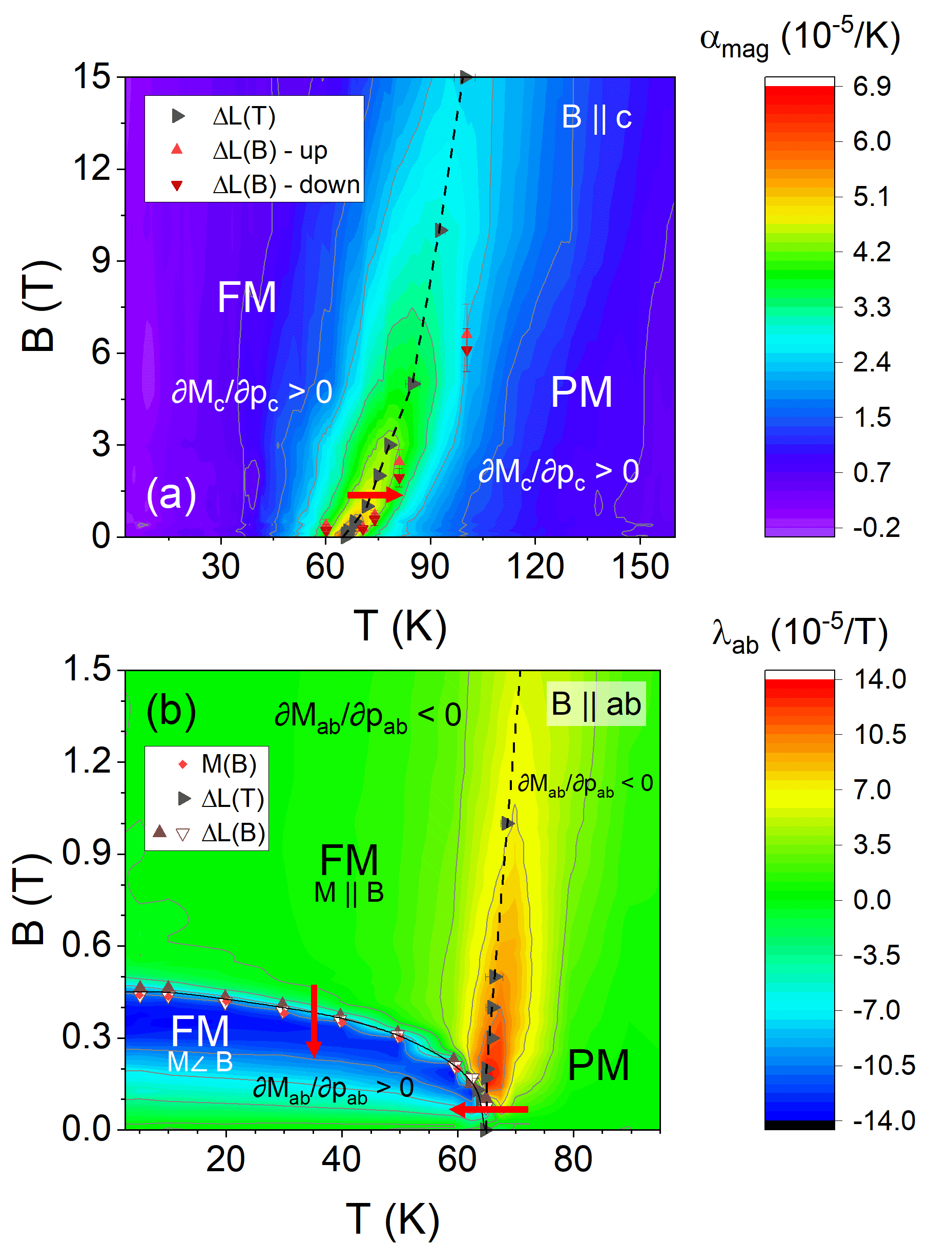}}
	\caption[] {\label{PD} Phase diagrams for $B\parallel c$ (a) and $B\parallel ab$ (b) with uniaxial pressure dependencies of the magnetization (see text) as well as of the phase boundaries (red arrows) indicated. Color coding shows the magnetic thermal expansion coefficient (a) as well as the magnetostriction coefficient (b). Dashed lines indicate the crossover from the FM to the PM phase. The solid black line in (b) indicates the transition from the low temperature and field (LTF) phase to the phase where magnetic moments are fully aligned along the applied field. Red arrows show the effects of uniaxial pressure (see the text).}
\end{figure}

From the thermal expansion, magnetostriction, and magnetization measurements, the phase diagrams for $B \parallel c$ and $B \parallel ab$ are derived (Fig.~\ref{PD}).
Dashed lines indicate the crossover from the ferromagnetic to the paramagnetic phase. Red arrows show the effects of uniaxial pressure, i.e., the stabilization of the FM low-temperature phase for $B,p \parallel c$, and strong suppression of the LTF phase for $B,p \parallel ab$. 
Additionally, the sign of the pressure dependence of the magnetization as calculated above is indicated in each of the phases.

Overall, from the anomalies in the strain response functions to 'magnetoelastic' phase diagram is consructed as they separate the different phases recently reported in \CGT , i.e., the paramagnetic (PM), ferromagnetic low-temperature/low-field (LTF) and aligned ferromagnetic (FM) phases for $B\parallel ab$ and PM/FM phases for $B\parallel c$~\cite{Zeisner2019, Selter2020}. The elastic signatures of the crossover between the PM to the FM phase allows to follow this feature up to high magnetic fields; it displays a left-bending behavior and a moderate field dependence. In addition, anomalous contributions to the thermal expansion and magnetostriction confirm the presence of short-range spin correlations well above \tc . As illustrated by the blue region in Fig.~\ref{PD}(b), full spin alignment driven by magnetic fields $B\|ab$ is associated with a jump in the magnetostriction coefficient, i.e., a kink in the magnetostrictive length changes, which signals a continuous phase transition. Very strong uniaxial pressure dependencies of the magnetization  in the LTF phase and of $B_{\mathrm{cr}}$ show the relevance of critical behavior in the whole LTF phase. In the vicinity of \tc , the presence of two distinct thermal expansion anomalies (at small applied magnetic field $B\parallel ab$) as well as the suppression of the transition temperature into the LTF phase in magnetic fields clearly supports the scenario of an intermediate ferromagnetic phase, i.e., the fully aligned FM phase, separating the LTF and PM phases at small applied magnetic fields~\cite{Selter2020}. Our data suggest that the anomalies associated with the crossover FM/PM and the LTF/FM transition, respectively, merge into a tricritical point~\cite{Liu2017}. In the vicinity of the tricritical point, we observe significantly enhanced spin fluctuations as indicated by strong magnetostrictive reponse (red and blue regions in Fig.~\ref{PD}(b)).  

\section{Summary}
Thermal expansion and magnetostriction measurements of \CGT\ single crystals confirm the presence of a low-temperature and low-field (LTF) phase for $B\parallel ab$. Magnetostriction and magnetization data of this phase can be modeled within the Stoner-Wohlfarth model at low temperatures, advocating that as a magnetic field $B\parallel ab$ is applied the magnetic moments rotate from the $c$ axis towards the applied field until saturation is reached around 0.5~T.
The uniaxial pressure dependence of the critical field shows that pressure $p\parallel ab$ strongly suppresses the LTF phase, presumably by lowering the uniaxial anisotropy. This conclusion is further supported by the uniaxial pressure dependence of the magnetization. In the vicinity of \TC , the presence of two distinct thermal expansion anomalies at small fields $B\parallel ab$ of different magnetic field dependence clearly supports the scenario of an intermediate FM phase separating PM and LTF phases at small fields and is indicative of a tricritical point. The observed very strong uniaxial pressure dependencies in particular in the LTF phase and in the vicinity of the tricritical point as well as pronounced critical behavior observed in the thermal expansion and the magnetostriction underlines the importance of spin fluctuations in \CGT\ and evidences the strong coupling of spin and lattice. 



\begin{acknowledgements}
We acknowledge financial support by BMBF via the project SpinFun (13XP5088) and by Deutsche Forschungsgemeinschaft (DFG) under Germany’s Excellence Strategy EXC2181/1-390900948 (the Heidelberg STRUCTURES Excellence Cluster) and through Projects No. KL 1824/13-1 (R.K.) and No. AS 523/4-1 (S.A.). B.B. acknowledges the W\"{u}rzburg-Dresden Cluster of Excellence on Complexity and Topology in Quantum Matter – ct.qmat (EXC 2147, Project No. 390858490).
\end{acknowledgements}

\bibliography{Cr2Ge2Te6_Paper_bibliography}


\end{document}


\title{Supplemental Material}
    \author{S. Spachmann$^{1,}$\footnote{sven.spachmann@alumni.uni-heidelberg.de}, , S. Selter$^{2}$, B. B\"uchner$^{2,3}$, S. Aswartham$^{2,}$\footnote{s.aswartham@ifw-dresden.de}, R. Klingeler$^{1,}$\footnote{klingeler@kip.uni-heidelberg.de}}

	\affiliation{$^1$Kirchhoff Institute for Physics, Heidelberg University, Heidelberg, Germany}
	\affiliation{$^2$Leibniz Institute for Solid State and Materials Research (IFW), Dresden, Germany}
	\affiliation{$^3$Würzburg-Dresden Cluster of Excellence ct.qmat}

\date{\today}
\pacs{} \maketitle

\noindent In the following 
\begin{itemize}
    \item a summary of anomalies in the magnetostriction and magnetization measurements for $B\parallel ab$ and resulting uniaxial pressure dependencies,
    \item the demagnetization factor and sample shape correction of magnetization measurements, 
    \item the thermal expansion coefficient around \TC\ at low fields up to 0.2~T, as well as
    \item magnetostriction and magnetization data up to 15~T
\end{itemize}
are shown which support the results of our manuscript.

\section{Pressure dependence of the critical field}
A summary of the relevant anomalies obtained by dilatometry and magnetometry at the phase boundary of the low temperature and low field (LTF) phase ($B\parallel ab$) as well as the derived pressure dependencies of the critical field $B_c$ are given in Table~\ref{Tab_dBdp}.

\renewcommand{\arraystretch}{1.2}
\begin{table}[!ht]
\setlength{\tabcolsep}{4pt}
    \centering
    \caption{Summary of anomalies in the magnetostriction and magnetization measurements for $B\parallel ab$ and derived uniaxial pressure dependence of the critical field.}
    \label{Tab_dBdp}
    \begin{tabular}{ccccc}
    \hline \hline
        $T$ & $B_c$ & $\Delta\lambda$ & $\Delta$(${\partial}M/{\partial}B)$ & ${\partial}B_c/{\partial}p_{ab}$ \\
        (K) & (T) & ($10^{-5}$ T$^{-1}$) & ($\mu_B $T$^{-1} $Cr$^{-1}$) & (T/GPa) \\  \hline
        2.25 & 0.46 & 19(3) & -4.6(5) & -1.25(18) \\ 
        5.1 & 0.46 & 19(2) & -4.6(5) & -1.22(18) \\
        10 & 0.46 & 19(2) & -4.5(5) & -1.26(18) \\ 
        19.8 & 0.43 & 19(2) & -4.5(5) & -1.24(18) \\ 
        29.7 & 0.406 & 18(2) & -4.7(6) & -1.19(17) \\
        39.7 & 0.37 & 18(2) & -4.8(6) & -1.11(16) \\ 
        49.6 & 0.315 & 19(2) & -5.2(7) & -1.07(16) \\
        59.3 & 0.226 & 24(2) & -5.4(7) & -1.33(19) \\ \hline \hline
    \end{tabular}
\end{table}
\renewcommand{\arraystretch}{1}

\clearpage 
\section{Corrections for Magnetization Measurements}

Measurements of the isothermal magnetization were performed on a thin sample with dimensions of $2.0\times 1.3\times 0.235$~mm$^3$ and a mass of $m = 3.27(5)$~mg. The measured magnetization was corrected for the demagnetizing field using demagnetization factors of $N = 0.746$ for $B\parallel c$ and $N = 0.099$ for $B\parallel ab$ (Fig.~\ref{SI_DemagCorr}(a) and (b)). Further correction factors due to the sample geometry were obtained based on the experimental data of a square nickel film \cite{QDNote}. To extract correction factors for the side lengths of 2.0~mm and 1.3~mm a square of equal area with a side length of 1.65~mm was assumed. For measurements with $B\parallel c$, i.e., along the thin direction, power-law fits to the data provided in Ref.~\cite{QDNote} for vibration amplitudes of 1.0~mm to 5.0~mm yielded the correction factor $f$ at 1.65 mm at different amplitudes. A linear interpolation of these data points resulted in $f = 1.086$ for the amplitude of 1.8~mm at which all measurements were performed.
For a vertical alignment of the sample, with $l_{ab} = 1.3$~mm and $B\parallel ab$, third-order
polynomial fits were applied to extract $f$ at different vibration amplitudes. A linear interpolation of the resulting values for the different amplitudes yielded $f = 1.061$ for an amplitude of 1.8~mm. The effect of the corrections is shown in Fig.~\ref{SI_DemagCorr}(d).

\renewcommand{\thefigure}{S1}
\begin{figure*}[ht]
	\center{\includegraphics [width=0.95\columnwidth,clip]{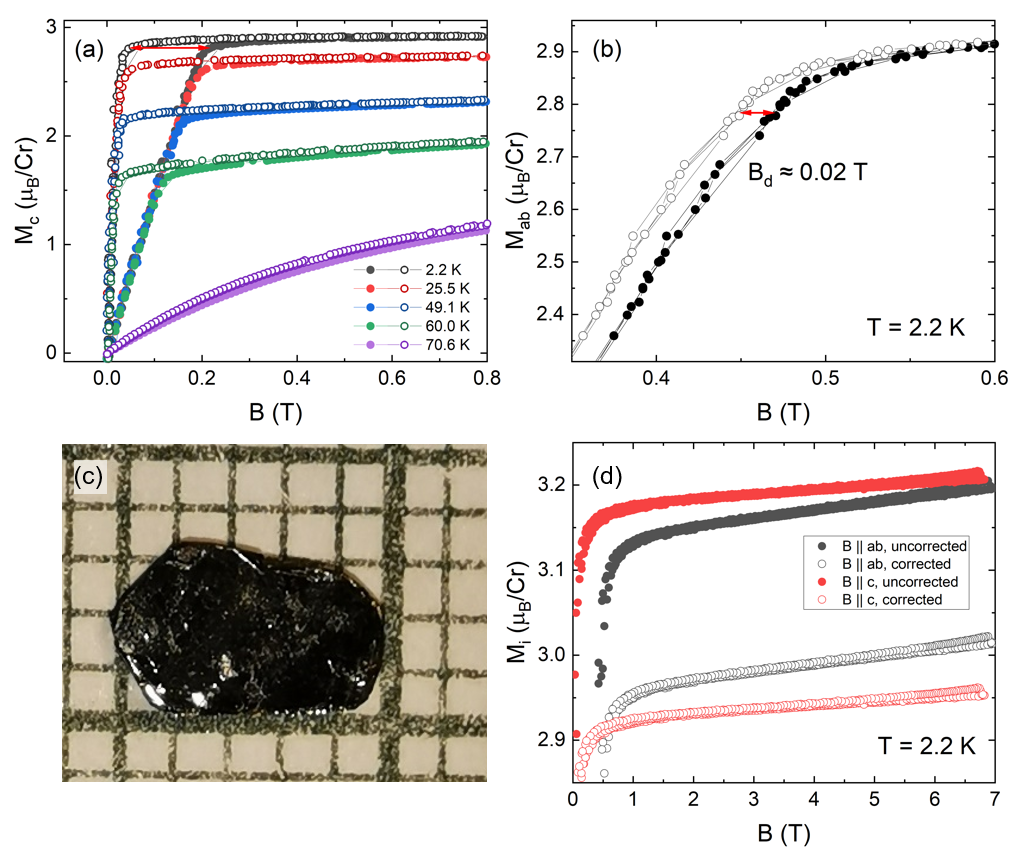}}
	\caption[] {\label{SI_DemagCorr} (a, b) Demagnetization factor correction with (a) $N = 0.746$ for $B\parallel c$ and (b) $N = 0.099$ for $B\parallel ab$. Red horizontal arrows give an impression of the demagnetization field. (c) \CGT\ sample used for the $c$ axis thermal expansion measurements. (d) Example for the effect of the magnetization correction due to sample geometry. Closed (open) symbols mark the uncorrected (corrected) data in (a), (b) and (d).}
\end{figure*}

\clearpage

\section{Further Magnetostriction and Magnetization data}

\renewcommand{\thefigure}{S2}
\begin{figure}[h]
	\center{\includegraphics [width=0.9\columnwidth,clip]{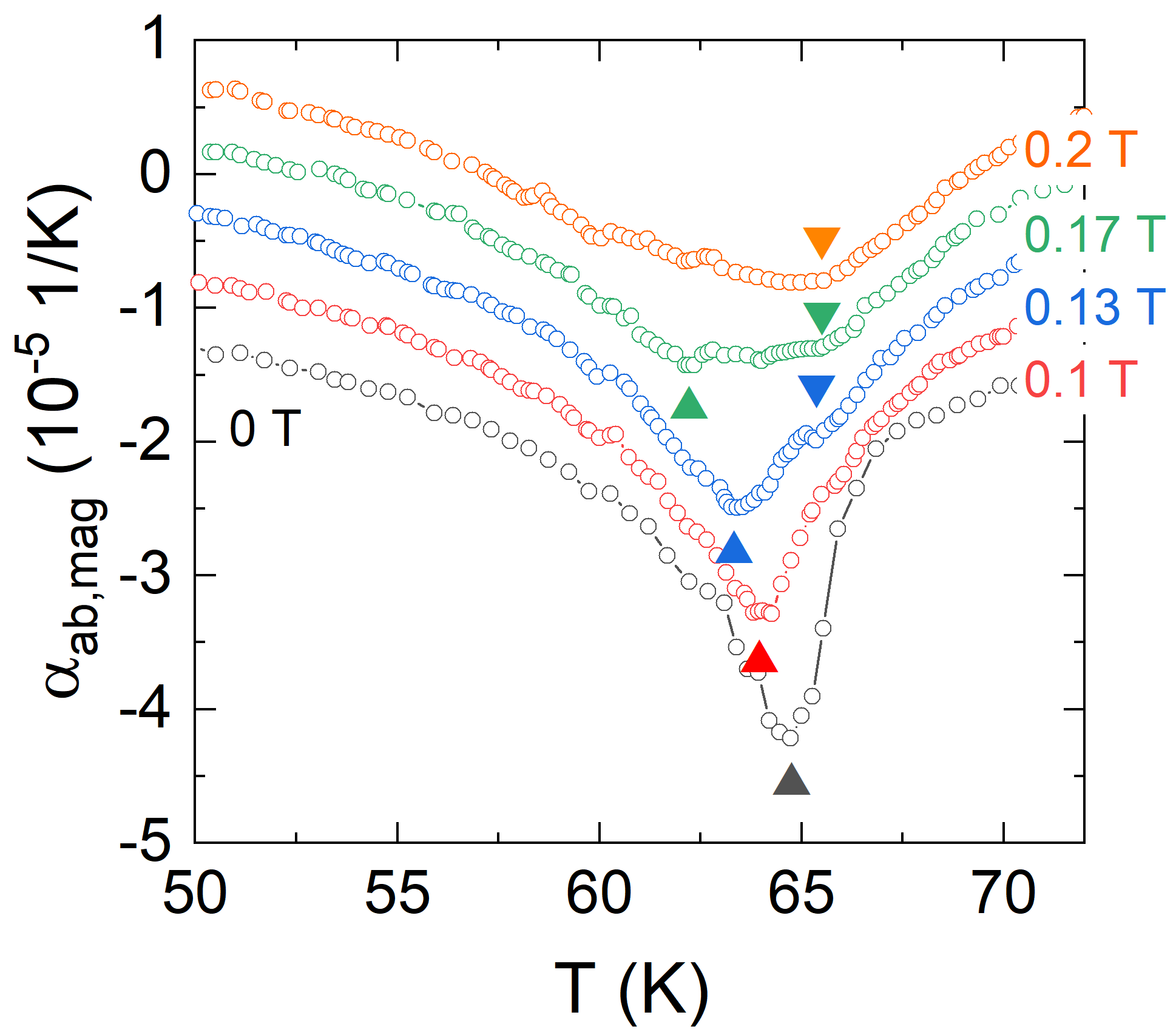}}
	\caption[] {\label{tri} Magnetic contributions to the thermal expansion coefficient in external fields $B\parallel ab \leq 0.2$~T (i.e., further enlargement of Fig.~1d in the main manuscript). Curves are offset by $5\times 10^{-6}$/K for clarity. Triangles indicate trends of anomalies. The data show the initial suppression of the main peak, two distinct anomalies at 0.13~T, and non-symmetric broadening of the upper peak as well as shift the anomaly onset to higher temperatures at 0.2~T.}
\end{figure}

\clearpage

A quadratic fit to the relative length changes ${\Delta}L_{ab}(B)$ in the LTF phase is shown in Fig.~\ref{Moment-vs-a}(a).
Fig.~\ref{Moment-vs-a} shows a comparison of the temperature dependence of the refined magnetic moment of Cr$^{3+}$ ions~\cite{Carteaux1995} to the proportionality constant $a(T)$ for the quadratic-in-field magnetostriction at different temperatures below \TC.
Fits up to fourth order of the relative length changes with respect to magnetization below $B_c$, ${\Delta}L_{ab}(M,T) = A_{1}(T)M^2+A_{2}(T)M^4$, are shown in Fig.~\ref{Moment-vs-a}(c).

\renewcommand{\thefigure}{S3}
\begin{figure}[ht]
	\center{\includegraphics [width=0.45\columnwidth,clip]{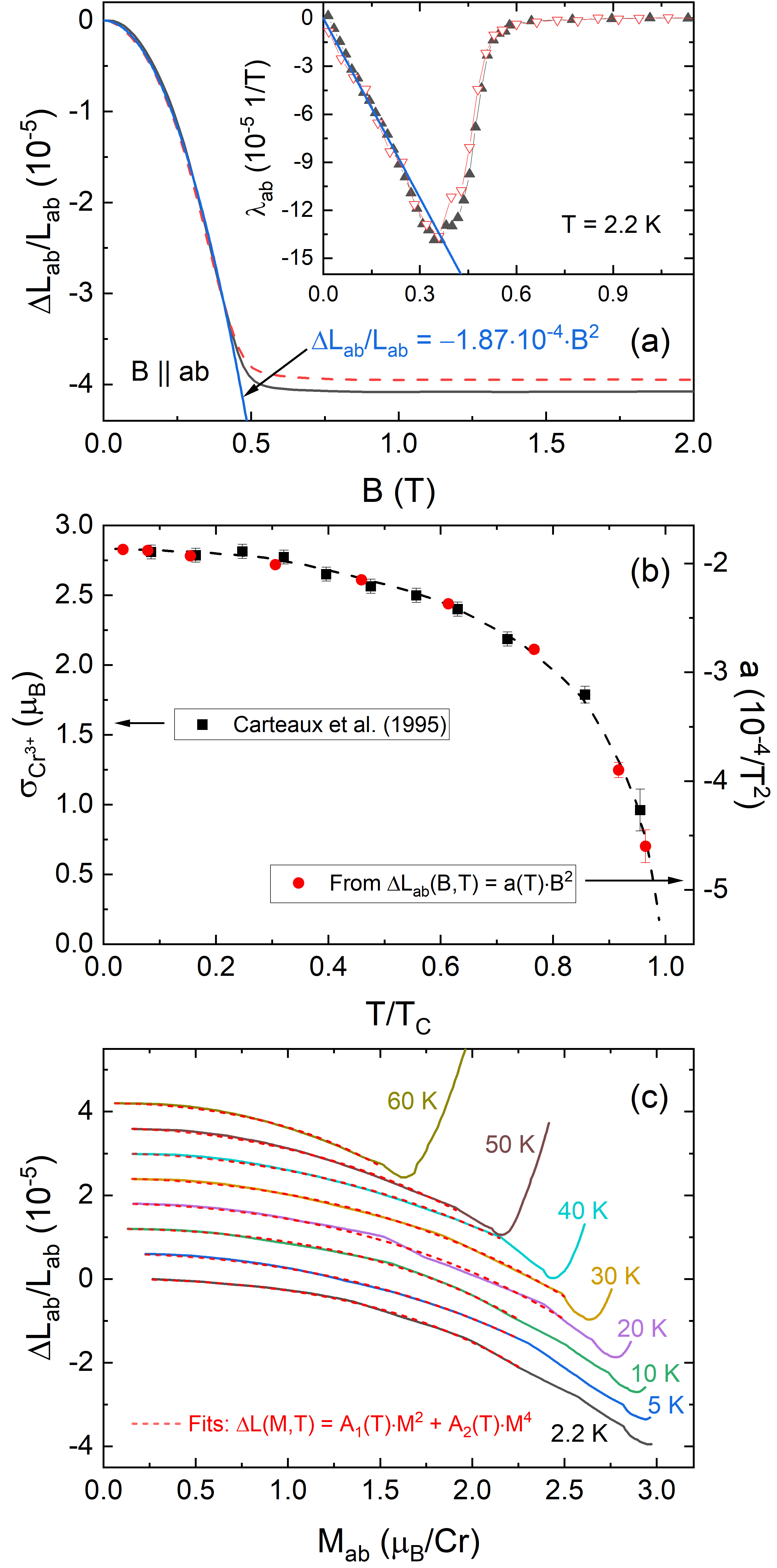}}
	\caption[] {\label{Moment-vs-a} (a) Magnetostriction of the in-plane direction ($B\parallel ab$) at $T \approx 2~K$. The inset shows the corresponding magnetostriction coefficient. The black line and triangles mark the up-sweep, the dashed red line and empty red triangles the down-sweep. The blue curve indicates a fit to a quadratic-in-field behavior. (b) Refined moment of Cr$^{3+}$ ions (left axis, digitized from Ref.~\onlinecite{Carteaux1995}) vs. proportionality factor $a(T)$ for quadratic-in-field magnetostriction ${\Delta}L_{\mathrm{ab}}/L_{\mathrm{ab}}$. The dashed line is a guide to the eye taken also from Ref.~\onlinecite{Carteaux1995}. (c) In-plane relative length changes as a function of in-plane magnetization. Data are shifted vertically by $6\cdot 10^{-6}$ for better visibility. Red dashed lines indicate fits ${\Delta}L= A_{1}(T)\cdot M_{\mathrm{ab}}^2+A_{2}(T)\cdot M_{\mathrm{ab}}^4$.}
\end{figure}

\clearpage

Fig.~\ref{SI_MSMB_c-axis} shows the magnetostriction and magnetostriction coefficients for $B\parallel c$ at various temperatures below \TC\ (a, b) as well as magnetization measurements at temperatures from 2 to 204~K (c, d).

\renewcommand{\thefigure}{S4}
\begin{figure*}[ht]
	\center{\includegraphics [width=0.95\columnwidth,clip]{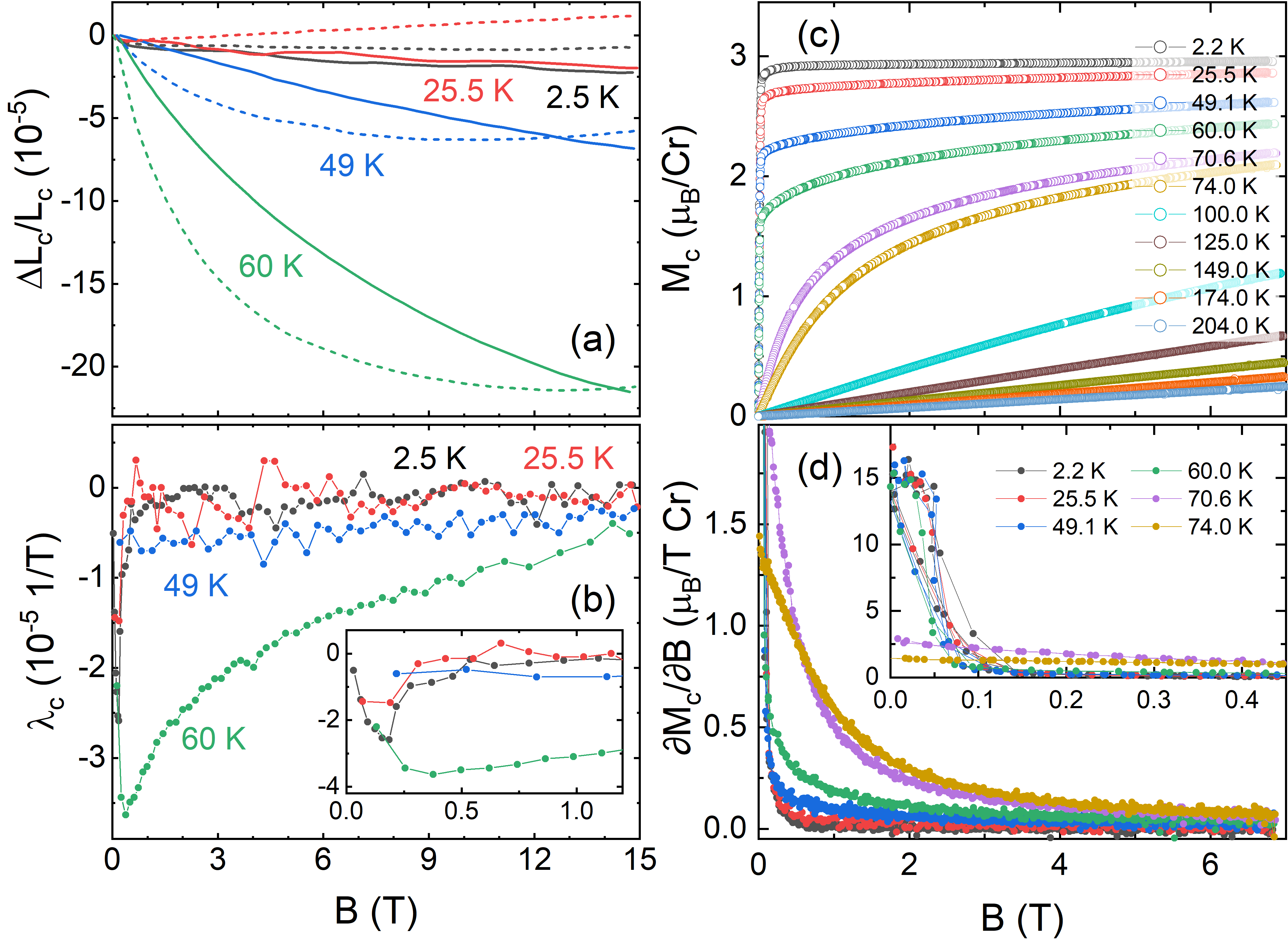}}
	\caption[] {\label{SI_MSMB_c-axis} Comparison of magnetostriction (a, b) and magnetization (c, d) for $B\parallel c$. (a) Relative length changes and (b) magnetostriction coefficient for temperatures from 2~K to 60~K. (c) Isothermal magnetization and (d) magnetic susceptibility at different temperatures up to 204~K. Magnetization data was corrected for demagnetization effects whereas magnetostriction data is shown as measured.}
\end{figure*}

\clearpage

Magnetostriction and magnetostriction coefficients for $B\parallel ab$ at various temperatures below \TC\ as well as magnetization measurements at temperatures from 2 to 199~K are shown in Fig.~\ref{SI_MSMB_In-plane}.

\renewcommand{\thefigure}{S5}
\begin{figure*}[ht]
	\center{\includegraphics [width=0.95\columnwidth,clip]{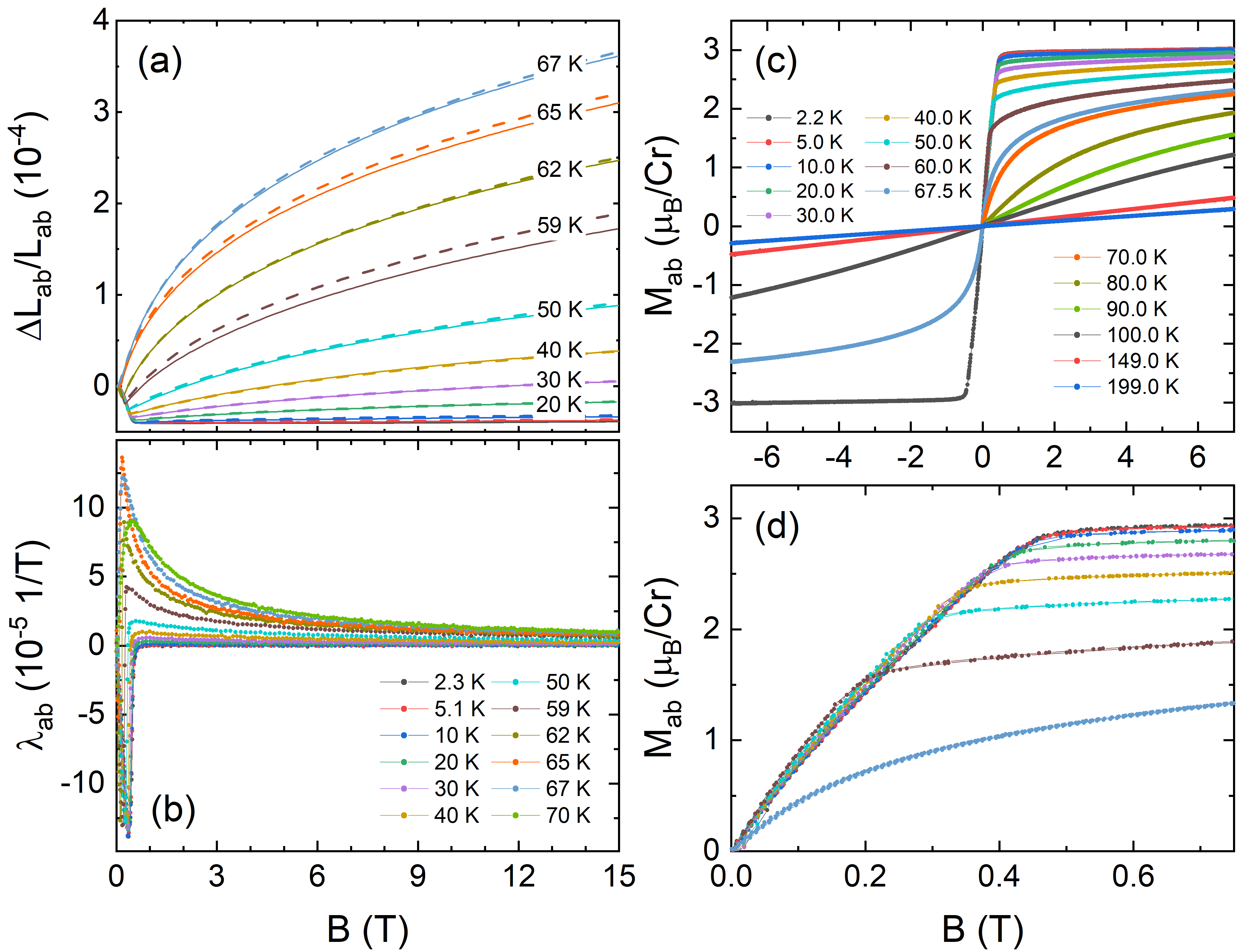}}
	\caption[] {\label{SI_MSMB_In-plane} Comparison of magnetostriction (a, b) and magnetization (c, d) for $B\parallel ab$. (a) Relative length changes and (b) magnetostriction coefficient for temperatures from 2~K to 70~K. (c) Isothermal magnetization, and (d) magnetic susceptibility at different temperatures up to 199~K. Magnetization data was corrected for demagnetization effects whereas magnetostriction data is shown as measured.}
\end{figure*}

\clearpage

Magnetostriction measurements of $B\parallel ab, c$ above \TC\ are shown in Fig.~\ref{SI_MS_HighT}. Both up- and down-sweeps are shown. Up-sweeps are taken from Ref.~\cite{SpachmannCGT2022}. While hysteresis is negligible for $B\parallel ab$, significant hysteresis is visible for $B\parallel c$ between 70 and 200 K.

\renewcommand{\thefigure}{S6}
\begin{figure*}[htbp]
	\center{\includegraphics [width=0.95\columnwidth,clip]{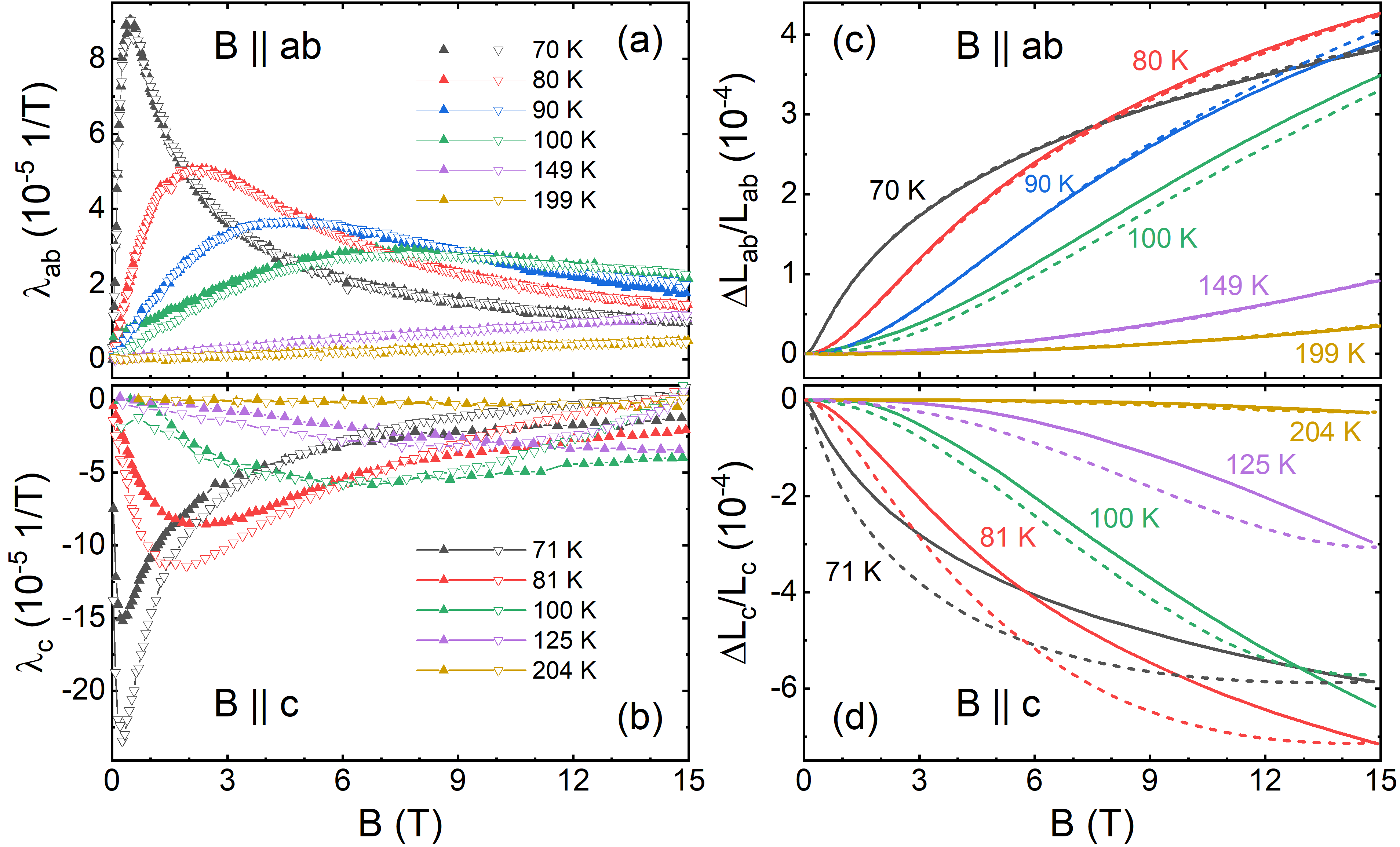}}
	\caption[] {\label{SI_MS_HighT} (a, b) Magnetostriction coefficients and (c, d) magnetostrictive relative length changes for $B\parallel ab$ (a, c) and $B\parallel c$ (b, d) at temperatures $T > T_{\mathrm{C}}$. Filled upward triangles and solid lines mark up-sweeps, empty downward-triangles and dashed lines mark down-sweeps. Up-sweeps are taken from Ref.~\cite{SpachmannCGT2022}.}
\end{figure*}

\bibliography{Cr2Ge2Te6_Paper_bibliography}